\documentclass[11pt,twoside]{article}
\usepackage{./asp2014}

\aspSuppressVolSlug
\resetcounters

\begin{document}

\title{The VO: A powerful tool for global astronomy}
\author{Christophe~Arviset$^1$, Mark~Allen$^2$, Alessandra~Aloisi$^3$, Bruce~Berriman$^4$, Catherine~Boisson$^5$, Baptiste~Cecconi$^6$, David~Ciardi$^4$, Janet~Evans$^7$, Giuseppina~Fabbiano$^7$, Francoise~Genova$^2$, Tim~Jenness$^8$, Bob~Mann$^9$, Tom~McGlynn$^{10}$, William~OMullane$^{11}$, David~Schade$^{12}$, Felix~Stoehr$^{13}$, and Andrea~Zacchi$^{14}$}
\affil{$^1$ESAC Science Data Centre, ESA, Spain; \email{Christophe.Arviset@esa.int}}
\affil{$^2$CDS, Universit\'e de Strasbourg, CNRS, France }
\affil{$^3$MAST, STScI, USA}
\affil{$^4$IPAC, Caltech, USA}
\affil{$^5$LUTH, Observatoire de Paris, CNRS, France}
\affil{$^6$LESIA, Observatoire de Paris, CNRS, France}
\affil{$^7$SAO, CXC, USA}
\affil{$^8$LSST Project Management Office, Tucson, AZ, USA}
\affil{$^9$Wide-Field Astronomy Unit, University of Edinburgh, UK}
\affil{$^{10}$NASA, HEASARC, USA}
\affil{$^{11}$Gaia, ESAC, ESA, Spain}
\affil{$^{12}$CADC, Canada}
\affil{$^{13}$ALMA, Germany}
\affil{$^{14}$INAF-OATs, Euclid, Italy}

\paperauthor{Christophe~Arviset}{Christophe.Arviset@esa.int}{}{European Space Agency}{ESAC Science Data Centre}{Villanueva de la Canada}{Madrid}{28691}{Spain} 
\paperauthor{M. G. Allen}{mark.allen@astro.unistra.fr}{0000-0003-2168-0087}{Observatoire astronomique de Strasbourg, Université de Strasbourg, CNRS, UMR 7550}{}{Strasbourg}{}{67000}{France}
\paperauthor{Alessandra~Aloisi}{aloisi@stsci.edu}{}{STScI}{Data Management}{Baltimore}{MD}{21218}{USA}
\paperauthor{G. Bruce~Berriman}{gbb@ipac.caltech.edu}{0000-0001-8388-534X}{California Institute of Technology}{IPAC}{Pasadena}{CA}{91125}{USA}
\paperauthor{Catherine~Boisson}{catherine.boisson@obspm.fr}{}{Observatoire de Paris, PSL Research University, CNRS, Univ. Paris Diderot, Sorbonne Paris Cit\'e}{LUTH}{Meudon}{}{92195}{France}
\paperauthor{Baptiste~Cecconi}{baptiste.cecconi@obspm.fr}{0000-0001-7915-5571}{Observatoire de Paris, PSL Research University, CNRS, Sorbonne Universit\'es, UPMC Univ. Paris 06, Univ. Paris Diderot, Sorbonne Paris Cit\'e}{LESIA}{Meudon}{}{92195}{France}
\paperauthor{David~R.~Ciardi}{ciardi@ipac.caltech.edu}{0000-0002-5741-3047}{IPAC/Caltech}{}{Pasadena}{CA}{91125}{USA}
\paperauthor{Janet~D.~Evans}{janet@cfa.harvard.edu}{0000-0003-3509-0870}{Smithsonian Astrophysical Observatory}{HEAD/CXC}{Cambridge}{MA}{02138}{USA}
\paperauthor{G. Fabbiano}{gfabbiano@cfa.harvard.edu}{0000-0002-3554-3318}{Smithsonian Astrophysical Observatory}{HEAD/CXC}{Cambridge}{MA}{02138}{USA}
\paperauthor{Francoise Genova}{francoise.genova@astro.unistra.fr}{0000-0002-6318-5028}{Observatoire astronomique de Strasbourg, Université de Strasbourg, CNRS, UMR 7550}{Observatoire Astronomique de Strasbourg}{Strasbourg}{N/A}{F-67000}{France} 
\paperauthor{Tim~Jenness}{tjenness@lsst.org}{0000-0001-5982-167X}{LSST}{Data Management}{Tucson}{AZ}{85719}{USA}
\paperauthor{Robert~G.~Mann}{rgm@roe.ac.uk}{0000-0002-0194-325X}{University of Edinburgh}{Institute for Astronomy}{Edinburgh}{Midlothian}{EH9 3HJ}{United Kingdom}
\paperauthor{Tom McGlynn}{tom.mcglynn@nasa.gov}{0000-0003-3973-432X}{NASA/GSFC}{HEASARC}{Greenbelt}{MD}{21228}{USA} 
\paperauthor{William~OMullane}{William.OMullane@sciops.esa.int}{0000-0003-4141-6195}{European Space Agency}{ESA Space Astronomy Centre}{Villanueva de la Canada}{Madrid}{28691}{Spain}
\paperauthor{David Schade}{David.Schade@nrc.ca}{0000-0002-4677-1586}{National Research Council Canada}{Canadian Astronomy 
Data Centre}{Victoria}{British Columbia}{V9B1K9}{Canada}
\paperauthor{Felix~Stoehr}{fstoehr [at] eso [dot] org}{}{ALMA}{ALMA Regional Centre}{Garching}{}{85748}{Germany}
\paperauthor{A.~Zacchei}{zacchei@oats.inaf.it}{0000-0003-0396-1192}{INAF-OATs}{}{Trieste}{}{I-34143}{Italy}

\begin{abstract}
Since its inception in the early 2000's, the Virtual Observatory (VO), developed as a collaboration of many national and international projects, has become a major factor in the discovery and dissemination of astronomical information worldwide. The International Virtual Observatory Alliance (IVOA) has been coordinating all these efforts worldwide to ensure a common VO framework that enables transparent access to and interoperability of astronomy resources (data and software) around the world.

The VO is not a magic solution to all astronomy data management challenges but it does bring useful solutions in many areas borne out by the fact that VO interfaces are broadly found in astronomy's major data centres and projects worldwide. Astronomy data centres have been building VO services on top of their existing data services to increase interoperability with other VO-compliant data resources to take advantage of the continuous and increasing development of VO applications. VO applications have made multi-instrument and multi-wavelength science, a difficult and fruitful part of astronomy, somewhat easier.

More recently, several major new astronomy projects have been directly adopting VO standards to build their data management infrastructure, giving birth to `VO built-in' archives. Embracing the VO framework from the beginning brings the double gain of not needing to reinvent the wheel and ensuring from the start interoperability with other astronomy VO resources. Some of the IVOA standards are also starting to be used by neighbour disciplines like planetary sciences.

There is still quite a lot to be done on the VO, in particular tackling the upcoming big data challenge and how to find interoperable solutions to the new data analysis paradigm of bringing and running the software close to the data.

We report on the current status and also desire to encourage others to adopt VO technology and engage them in the effort of developing the VO. Thus, we wish to ensure that the VO standards fit new astronomy projects requirements and needs.

\end{abstract}

\section{IVOA today and VO successes}
The International Virtual Observatory Alliance was created in 2002 and today consists of twenty diverse member projects worldwide. Over the years, some new national projects have joined, some others have withdrawn (some due to lack of funding) but most of them have persisted. Such international collaboration represents undoubtedly IVOA's greatest success, where people coming from all over the planet have managed to agree (sometimes with great difficulty) and implement astronomical data interoperability standards. 

The two annual IVOA interoperability meetings continue to be well attended with 70 to 100 participants, organised in working groups (Applications, Data Access Layer, Data Model, Grid and Web Services, Registry and Semantics) and interest groups (Data Curation and Preservation, Education, Theory, Time Domain, Operations, Knowledge Discovery in Databases).  The IVOA Executive Committee, with representatives from each VO project, makes decision in a collaborative manner, coordinates and oversees all IVOA activities.

After the first years of general brainstorming, getting the VO off the ground, the standard development process has now become very mature. Standards definition represents the main activity of the IVOA and in 2010, an effort was made to define more clearly and carefully the IVOA architecture \citep{2012ASPC..461..259A} which has remained very stable since then, with well established interoperability standards for tables, images, spectra and registries. IVOA work has always been done in an open and collaborative environment, with shareable software and infrastructure components, like registry validators, TAP libraries, VOTable parsers, data publishing software and tools (e.g. DaCHS).

On one side, many major astronomical data collections are now registered in and accessible through the VO, making them easily discoverable through VO portals (e.g. MAST, Datascope, ESASky). This was usually made by adding a VO layer on top of existing data archives, enabling interoperability amongst them. More recently, the VO protocols have been used to actually build astronomical data management infrastructures (such as the CADC, SkyMapper and the Gaia archive), converting these into the first `VO built-in' archives, which will benefit directly from interoperability with others VO compliant archives.

On the other side, more and more VO applications now enable end users to more easily access and combine data that comes from various data centres, regardless of how and where these products are being stored. The two most noticeable examples are Aladin and Topcat, which have now become widely used by astronomers.

Although specific VO funding has decreased or has been shifted over recent years, it is still recognised as an important e-infrastructure. In Europe, the VO is included in the ASTRONET European Infrastructure Roadmap, and the new ASTERICS project aims to make ESFRI (European Strategy Forum on Research Infrastructures) projects and their pathfinders data available for discovery. In the USA, VO support is provided by NASA data centres. The VO effort is becoming more an integral part of data centres and laboratory data management resources, rather than a side project. It is also noteworthy that the VO standards and protocols are being re-used by other scientific disciplines, like planetary science (EuroPlaNet), and for molecular and lines databases (VAMDC).

\section{VO problem areas and challenges}
Along with its successes, the VO is also experiencing some difficulties. Due to the decrease of VO direct funding in the UK, at ESO and more recently in the USA, the perception by some is that the VO is flickering out. Sometimes the VO is seen as a closed shop, where VO standards are complex and difficult to implement. Initial `simple' access protocols (Cone Search, Simple Image Access) were easier to implement while more effort is required for the more `rich' VO services (synchronous and asynchronous Table Access Protocol (TAP) with table upload and cross match functions, image cut outs, multi dimensional data, workflow building, storage of results on the server through VOSpace, etc.). These more sophisticated services are required to respond to the more complex data access and exploitation science cases. 

On the other hand, the VO initial expectations, as they were sometimes stated, were probably unrealistic. There was the need then to get it off the ground with great vision and ideas, but the VO was probably oversold. Many people might have had a wrong perception of the VO. The VO is not an astronomy `killer' application but rather a data management interoperability infrastructure. The IVOA role is to define the VO ecosystem and its interoperability standards and it is up to the astronomy projects and data centres to build VO services and VO applications. 

In the era of very large data and with the ever-growing need of the scientific community to connect datasets from different projects and archives, the VO can play a crucial role in helping the community access and utilize the data, but it still needs greater community take-up, in particular by the new projects such as LSST. The IVOA needs to find a way to better engage large data centres and projects, capturing their requirements in the standards development process, and trying to align their constraints and priorities with those of the IVOA (data cube, time domain, big data and bring the software to the data). Two VO implementation models can be envisaged: (1) VO layer, supported by VO publishing tools for small data centres with little IT expertise and, (2) VO built-in, supported by sets of VO software libraries for bigger data centres and projects where more expertise usually reside. The VO is being used more and more by data centres to build their infrastructure and this probably represents the future of the VO. Therefore, the IVOA needs to engage big projects and data centres as `participants', not as `customers'.

With an increasing number of VO services and VO built-in infrastructures, the VO has become operational. Hence, the IVOA needs now to play a more active role in ensuring that the VO ecosystem is reliable and trustworthy. Through the recently created IVOA Operations Group, this will be coordinated by curating the resources registered in the IVOA registries, monitoring their uptime and their compliance to standards, and encouraging the service providers to take appropriate measures to improve their services.

Measuring the success of the VO is another challenge for the IVOA. The VO is in use, but how can we see that? How do we define success (or failure) metrics? Counting the VO scientific publications probably won't capture the VO use, as it is unlikely that scientists acknowledge the VO infrastructure that enables them to do new science. Some may not even know that the services and software they use are based on VO protocols. Nobody acknowledges the web, although all scientists use it! Another option would be to try to define VO services usage statistics, but how can we uniformly collect and compare them? When astronomers use Topcat, MAST, Aladin, CADC services or ESASky, do they know they are using the VO? Do they need to know? Probably not. It could well be that the success of the VO is to be `invisible'!

The VO greatly facilitates data discovery and quick exploration of data from various data sources and data centres. Interoperability between SAMP compliant tools and archives allow easily to transfer data to dedicated external applications for display and analysis. This mechanism is in place and works well for tables and images, but still needs to be improved for spectra, multi dimensional and time domain data. The VO can not do the science for astronomers, but its goal is to make access and use of data much easier to enable new science that could not be easily done otherwise.

\section{Conclusions}
The VO is not a magic solution to all astronomy data management challenges but it offers some powerful tools and useful solutions that can help the scientific community tackle big data challenges it faces. IVOA standards can help building archive data centre infrastructure and they enable interoperability amongst datasets coming from various instruments and wavelengths, facilitating new science. The VO is not astronomy's ultimate application but VO applications will be key to its success.

If one wants to take something out of the IVOA, one needs to bring something in. The IVOA needs to better engage data centre and big projects to participate in standards development so they can better fit their data management needs. 

The VO is vibrant, spreading its use within data centres (e.g. ESAC, CADC, CDS, CfA, STScI) and big projects (e.g. Gaia, Euclid, ALMA, EuroPlaNet, CTA). An ever increasing number of  VO applications are being built to exploit the standard interfaces.  Now work must continue towards a reliable and operational VO infrastructure, invisible but extremely useful for the scientific community. The VO must become the fabric of astronomical research.


\end{document}